\documentclass[twocolumn,showpacs,preprintnumbers,amsmath,amssymb]{revtex4}

\usepackage{graphics}

\begin{document}

\title{Statistical theory of high-gain free-electron laser
saturation}

\author{
Julien Barr{\'e}$^{1,2,*}$\thanks{Julien.Barre@ens-lyon.fr,
$*$ Present address: Theoretical Division, Los Alamos National Laboratory, USA
},
Thierry Dauxois$^{1}$\thanks{Thierry.Dauxois@ens-lyon.fr},
Giovanni De Ninno$^{3}$\thanks{giovanni.deninno@elettra.trieste.it},
Duccio Fanelli$^{4}$\thanks{fanelli@et3.cmb.ki.se},
Stefano Ruffo$^{2}$\thanks{ruffo@ingfi1.ing.unifi.it}}

\affiliation{1. Laboratoire de Physique, UMR-CNRS 5672, ENS Lyon,
46
All\'{e}e d'Italie, 69364 Lyon c\'{e}dex 07, France\\
2. Dipartimento di Energetica
and CSDC, Universit{\`a} di Firenze, INFM and
INFN,
via S. Marta, 3, 50139 Firenze, Italy\\
3. Sincrotrone Trieste, Trieste, Italy\\
4. Department of Cell and Molecular Biology, Karolinska Institute,
SE-171 77 Stockholm, Sweden}

\date{\today}

\begin{abstract}
  We propose a new approach, based on statistical mechanics, to
  predict the saturated state of a single-pass, high-gain
  free-electron laser (FEL). In analogy with the violent relaxation
  process in self-gravitating systems and in the Euler equation of 2D
  turbulence, the initial relaxation of the laser can be described by
  the statistical mechanics of an associated Vlasov equation. The laser
  field intensity and the electron bunching parameter reach a
  quasi-stationary value which is well fitted by a Vlasov stationary
  state if the number of electrons $N$ is sufficiently large.  Finite
  $N$ effects (granularity) finally drive the system to
  Boltzmann-Gibbs statistical equilibrium, but this occurs on times
  that are unphysical (i.e. excessively long undulators).  All
  theoretical predictions are successfully tested by means of finite $N$
  numerical experiments.
\end{abstract}


\maketitle

The interaction of charged particles with electromagnetic fields is a
topic of paramount importance in a large variety of physical
phenomena, from plasma dynamics to astrophysical systems. On the
energy exchange between particles and field also relies the
possibility of generating coherent and tunable radiation sources, such
as Free-Electron Lasers (FELs). In this case a relativistic electron
beam propagating through a periodic magnetic field (produced by an
undulator) interacts with a co-propagating electromagnetic wave.
Lasing occurs because the undulator field and the radiation combine to
produce a beat wave that travels slower than the speed of light and
can be synchronized with electrons. Among different schemes,
single-pass, high-gain FELs are currently attracting growing
interests~\cite{physicsworld}, as they are promising sources of
powerful and coherent light in the UV and X
ranges~\cite{leu,DESY01,list}. In the high-gain regime, both the
light intensity and the longitudinal bunching of the electron beam
increase exponentially along the undulator, until they reach
saturation due to nonlinear effects. Understanding this saturation
process is important to estimate, and then {\em optimize}, the
performance and building costs of a FEL.

Theoretical analyses usually rely on {\em dynamical} methods in
combination with detailed, but rather complicated, numerical
simulations. In this paper, we propose a new approach, which is based
on {\em statistical mechanics}, to study the saturated state of a
high-gain single-pass FEL. We restrict our analysis to the
steady-state regime, which amounts to neglect the variation of the
electromagnetic wave within the electron pulse length (small
electrons-radiation slippage). However, it is important to stress that
because of its intrinsic flexibility, we believe that our statistical
approach will be applicable also to alternative
schemes, such as harmonic generation~\cite{armo1}.

The starting point of our study is the
Colson-Bonifacio model~\cite{Bonifacio90}.
Under the hypotheses of one dimensional motion
and monochromatic radiation, the steady-state
equations for the $j^{\text{th}}$ electron of the beam coupled to
radiation read:
\begin{eqnarray}
\frac{{\mathrm d}\theta_j}{{\mathrm d}\bar{z}} &=& p_j\quad, \label{eq:mvttheta}\\
\frac{{\mathrm d}p_j}{{\mathrm d}\bar{z}} &=&
-\mathbf{A}e^{i\theta_j}-\mathbf{A}^{\ast}e^{-i\theta_j}\quad,
\label{eq:mvtp}\\
\frac{{\mathrm d}\mathbf{A}}{{\mathrm d}\bar{z}} &=& i\delta\mathbf{A}
+\frac{1}{N}\sum_j e^{-i\theta_j}~\quad, \label{eq:mvtA}
\end{eqnarray}
where $N$ is the number of electrons in a single radiation wavelength
and $\bar{z}=2k_u \rho z \gamma_r^2/\langle \gamma \rangle_0^2$ is the rescaled longitudinal
coordinate, which plays the role of time.  Here,
$\rho=(a_{w}\omega_p/4ck_u)^{2/3}/\gamma_r$ is the so-called Pierce parameter,
$\gamma_r$ the resonant energy, $\langle \gamma \rangle_0$ the mean energy of the electrons
at the undulator's entrance, $k_u$ the wave vector of the undulator,
$\omega_p=(e^2n/m\varepsilon_0)^{1/2}$ the plasma frequency, $c$ the speed of light,
$e$ and $m$ respectively the charge and mass of one electron. Further,
$a_w=(eB_w/k_umc^2)$, where $B_w$ is the rms undulator field, for the
case of a helical undulator.  By introducing $k$ as
  the wavenumber of the FEL radiation, the phase $\theta$ is defined by
$\theta=(k+k_u)z-2\delta\rho k_u z \gamma_r^2/\langle \gamma \rangle_0^2$ and its conjugate momentum
$p=(\gamma-\langle \gamma \rangle_0)/(\rho \langle \gamma \rangle_0)$.  $\mathbf{A}$ is the scaled field
amplitude, a complex vector, transversal to $z$,
$\mathbf{A}=(A_x,A_y)$\cite{aciter}.  Finally, the detuning parameter
is given by $\delta=(\langle \gamma \rangle_0^2-\gamma_r^2)/(2\rho\gamma_r^2)$, and measures the average
relative deviation from the resonance condition.

Although very simple, such a model captures the main features of the
dynamics of the single-pass FEL, as shown by a systematic comparison
with numerical predictions based on more complete approaches
\cite{perseo,genesis}. Using this model, we are able to predict
analytically the mean saturated laser intensity, the electron beam
bunching, and the electrons' velocity distribution, for a wide class
of initial conditions (i.e. energy spread, bunching and radiation
intensity). The analytical results agree very well with numerical
simulations.

The above system of equations can be derived from the Hamiltonian
\begin{equation}
H=\sum_{j=1}^N\frac{p_j^2}{2} -N \delta I +2\sqrt{I}\sum_{j=1}^N
\sin(\theta_j-\varphi) \label{eq:Hamiltonien},
\end{equation}
where the intensity~$I$ and the
phase~$\varphi$ of the wave are related to $\mathbf{A}=A_x+iA_y=\sqrt{I}\
e^{-i\varphi}$. In addition to the ``energy'' $H$, the total momentum $P=\sum_j
p_j + N \mathbf{A}\mathbf{A}^{\ast}$ is also a conserved quantity.  Let
us note that one can always take $P=0$, upon a shift in the
detuning~\cite{ElskensEscande}; thus, we always suppose $P=0$ in the
following.

It is important to emphasize that
Hamiltonian~(\ref{eq:Hamiltonien}) models the interaction between
radiation and electrons. Hence, it describes a quite
universal phenomenon which is encountered in many branches of
physics. As an example, in the context of plasma theory, the
so-called plasma-wave Hamiltonian~(\ref{eq:Hamiltonien})
characterizes the self-consistent interaction between a Langmuir
wave and $N$ particles, after an appropriate redefinition of the
variables involved~\cite{ElskensEscande}. Establishing a formal
bridge between these two areas allows to recast in the context of
the single-pass LINAC FEL numerous results
originally derived in the framework of plasma physics. In
addition, Hamiltonian~(\ref{eq:Hamiltonien}) can be viewed as a
direct generalization of mean-field
models~\cite{Houches02,barrethesis,Yoshi}, which are widely
studied nowadays because of their intriguing features:
statistical ensemble inequivalence, negative specific heat,
dynamical stabilization of out-of-equilibrium structures.

\begin{figure}[t]
\resizebox{0.4\textwidth}{!}{\includegraphics{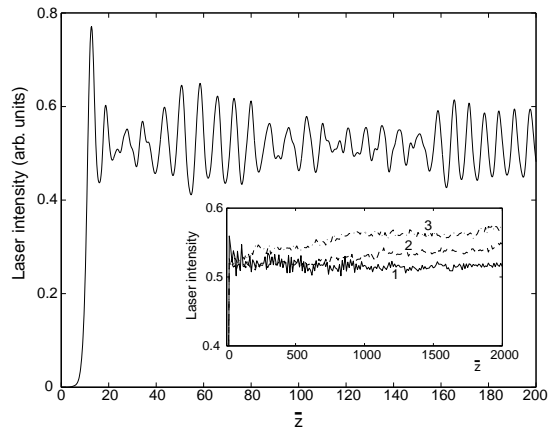}}
\caption{Typical evolution of the radiation
intensity using Eqs.~(\ref{eq:mvttheta})-(\ref{eq:mvtA}); the
detuning $\delta$ is set to $0$, the energy per electron $H/N=0.2$ and
$N=10^4$ electrons are simulated. The inset presents averaged
simulations on longer times for different values of $N$:
$5\cdot10^3$ (curve 1), 400 (curve~2) and 100 (curve~3).
\label{fig:saturation}}
\end{figure}

In plasma physics, it was numerically
shown~\cite{ElskensEscande,Firpo01} that, in the region of
instability, wave amplification occurs in two steps.
One first observes an exponential growth 
of the wave amplitude, followed by damped
oscillations around a well defined level. However, the
system does not reach a stationary state and this initial
stage is followed by a slow relaxation towards the final 
statistical equilibrium. An example of this behavior is 
shown in  Fig.~\ref{fig:saturation}.

The separation into two distinct timescales characterizes also the
dynamics of self-gravitating systems and is a well-known
phenomenon in astrophysics ~\cite{LyndenBell68,Chavanis96}.
The intermediate quasi-stationary states live longer and longer
as the number of particles $N$ is increased.
It is believed that galaxies ($N\simeq10^{11}$) are well described
by Vlasov equilibrium~\cite{LyndenBell68}, which characterizes the quasi-stationary
state of the $N$-particle system (see below). On the contrary,
Boltzmann-Gibbs statistics applies to the ``smaller" ($N\simeq10^6$)
globular clusters.

A typical evolution of the radiation intensity $I$ as a function
of the longitudinal coordinate $\bar{z}$,
according to the Free Electron Laser model
(\ref{eq:mvttheta})-(\ref{eq:mvtA}) is displayed
in Fig.~\ref{fig:saturation}: starting from a very weak radiation,
the intensity grows exponentially and saturates, oscillating
around a well defined value.
This growth and first relaxation of
the system (usually called''violent relaxation'' in astrophysics)
is governed by the Vlasov equation~\cite{Firpo98,barrethesis},
which is rigorously derived by taking the continuum limit ($N \to
\infty$ at fixed volume and energy per particle). 
On longer timescales, whose duration strongly depends on
the particles number~$N$ (see the inset of
Fig.~\ref{fig:saturation}), there is a slow drift of the intensity
of the beam towards the final asymptotic plateau determined by the
Boltzmann-Gibbs statistics. Such process is driven by granularity,
a finite-$N$ effect~\cite{LyndenBell68,Chavanis96,barrethesis}.
This final relaxation takes place on an extremely
long time scale, well beyond the physical constraints imposed by a
reasonable undulator length. We thus concentrate in the following
on the Vlasov description of the dynamics.

A linear analysis~\cite{Bonifacio90} leads directly to the
determination of the boundaries of the instability domain, which
are mainly controlled by the detuning $\delta$ and by the initial
energy per electron.  In the case of a monoenergetic electron
beam, the instability disappears for
$\delta>\delta_c\simeq 1.9$. Linear analysis also provides
estimates of the growth rate of $I$. However, getting insights on
the saturated state requires a nonlinear study of the
system; the standard approach to this problem is mainly dynamical,
as for instance in Ref.~\cite{Gluckstern93}. In the following we
discuss a new procedure, based on statistical mechanics.

As sketched in the previous discussion, we are interested
in the intermediate metastable state and, therefore, we will first consider the
statistical theory of the Vlasov equation, originally introduced in the
astrophysical context~\cite{LyndenBell68,Chavanis96}. The basic idea
is to coarse-grain the microscopic one-particle distribution function
$f(\theta,p,t)$, which is stirred and filamented at smaller and smaller
scales by the Vlasov time evolution. An entropy is then associated to the
coarse-grained distribution $\bar{f}$, which essentially counts the
number of microscopic configurations. Equilibrium is
then computed by maximizing this entropy while imposing the dynamical
constraints. A rigorous description of this procedure can be found
in Ref.~\cite{Michel94} in the context of two-dimensional Euler hydrodynamics.

In the continuum limit, Eqs.~(\ref{eq:mvttheta})-(\ref{eq:mvtA})
lead to the following Vlasov-wave system:
\begin{eqnarray}
\frac{\partial f}{\partial \bar{z}} &=& -p\frac{\partial
f}{\partial \theta}
+2(A_x\cos{\theta}-A_y\sin{\theta})\frac{\partial f}{\partial p}\quad ,
\label{eq:VlasovFELa}\\
\frac{\partial A_x}{\partial \bar{z}}   &=& -\delta
A_y+\frac{1}{2\pi} \int f \cos{\theta}  \, {\mathrm d}\theta \, {\mathrm d}p\quad ,
\label{eq:VlasovFELb}\\
\frac{\partial A_y} {\partial \bar{z}}  &=&  \delta
A_x-\frac{1}{2\pi} \int f \sin{\theta} \, {\mathrm d}\theta \, {\mathrm d}p\quad .
\label{eq:VlasovFELc}
\end{eqnarray}
Note that these equations have been studied numerically in a recent
work by Vinokurov \emph{et al.}~\cite{Kim01}, for the case
$\delta=0$. The Vlasov-wave
equations~(\ref{eq:VlasovFELa}-\ref{eq:VlasovFELc}) conserve the pseudo-energy
\begin{eqnarray}
\varepsilon =&& \int \!\!{\mathrm d}p{\mathrm d}\theta \,
\frac{p^2}{2}{f}(\theta,p) -\delta(A_x^2+A_y^2)\nonumber\\
 &&
+2\int \!\!{\mathrm d}p{\mathrm d}\theta \, {f}(\theta,p)
\left(A_x\sin \theta +A_y\cos \theta\right)\quad ,
\label{eq:energy}
\end{eqnarray}
and the total momentum
\begin{eqnarray}
\sigma &=& \int \!\!{\mathrm d}p{\mathrm d}\theta \ p{f}(\theta,p) +A_x^2+A_y^2~.
\label{eq:momentum}
\end{eqnarray}
For the sake of simplicity, let us suppose that the beam is initially
unbunched, and that energies are distributed according to a step
function, such that
\begin{eqnarray}
f(\theta,p,t=0) &=&f_0=
\frac{1}{4\pi \bar{p}} \qquad \text{if}\ -\bar{p}\leq p\leq \bar{p} \nonumber\\
                &=& 0 \quad\qquad\qquad\ \text{otherwise.}
\label{eq:distributionWB}
\end{eqnarray}
As far as one is dealing with small energy dispersions, the
profile~(\ref{eq:distributionWB}),  called waterbag initial
condition, represents a good approximation of a more natural
Gaussian initial distribution. Numerical tests fully confirm the
validity of this simple observation. According
to~(\ref{eq:distributionWB}), $f$ takes only two distinct values,
and coarse-graining amounts to perform a local average of both.
The entropy per particle associated with the
coarse-grained distribution $\bar{f}$ is then a mixing
entropy~\cite{Chavanis96,barrethesis} and reads
\begin{equation}
s(\bar{f})=-\int \!\!{\mathrm d}p{\mathrm d}\theta \, \left(\frac{\bar{f}}{f_0} \ln \frac{\bar{f}}{f_0}
+\left(1-\frac{\bar{f}}{f_0}\right)\ln
\left(1-\frac{\bar{f}}{f_0}\right)\right).
\label{eq:entropieVlasov}
\end{equation}
As the electromagnetic radiation represents only two degrees of
freedom within the $(2N+2)$ of Hamiltonian~(\ref{eq:Hamiltonien}),
its contribution to entropy can be neglected.

 The equilibrium
state is computed~\cite{barrethesis} by solving the constrained
variational problem:
\begin{eqnarray}
S(\varepsilon,\sigma) = \max_{\bar{f},A_x,A_y} \biggl(  s(\bar{f})
&&\biggr|
 H(\bar{f},A_x,A_y)=N\varepsilon;\; \int \!\! {\mathrm d}\theta {\mathrm d}p \bar{f}=1;\, \nonumber\\
&& \,P(\bar{f},A_x,A_y)=\sigma\biggr).
\label{eq:problemevar}
\end{eqnarray}
Introducing three Lagrange multipliers $\beta$, $\lambda$ and
$\mu$ for the energy, momentum and normalization constraints and
differentiating Eq.~(\ref{eq:problemevar}) with respect to $\bar{f}$,
one gets the equilibrium distribution
\begin{equation}
\label{eq:barf} \bar{f}=
f_0\frac{e^{-\beta(p^2/2+2A\sin\theta)-\lambda p-\mu}}
{1+e^{-\beta(p^2/2+2A\sin\theta)-\lambda p-\mu}}~.
\end{equation}
By differentiating Eq.~(\ref{eq:problemevar}) with respect to
$A_x$ and $A_y$, one obtains in addition the expression for the
amplitude of the wave
\begin{equation}
\label{eq:A} A=\sqrt{A_x^2+A_y^2}=\frac{\beta}{\beta\delta-\lambda}\int
\!\!{\mathrm d}p{\mathrm d}\theta\, \sin \theta \bar{f}(\theta,p).
\end{equation}
Using the above equations for the three constraints, the
statistical equilibrium calculation is now reduced to finding the
values of $\beta$, $\lambda$ and $\mu$ as functions of 
energy~$\varepsilon$ and total momentum~$\sigma$. This last step,
performed numerically using for example a Newton-Raphson method,
leads directly to the estimates of the main physical parameters.

Furthermore, let us stress that in the limit of a vanishing energy
dispersion, the area occupied by the $f=f_0$ level in the
one-particle phase space is small, so that the coarse-grained
distribution $\bar{f}$ verifies $\bar{f} \ll f_0$ everywhere. The
second term in the entropy~(\ref{eq:entropieVlasov}) is thus
negligible, and~(\ref{eq:barf}) reduces to the Gibbs distribution
\begin{equation}
\label{eq:barf2} \bar{f}\propto
e^{-\beta(p^2/2+2A\sin\theta)-\lambda p}~.
\end{equation}
Vlasov equilibrium is in that case equivalent to the full
statistical equilibrium.
Then, solving the constraint equations yields
\begin{eqnarray}
b &=&
A^3-\delta A
\label{eq:eqstatm} \\
 A &=& \left( \varepsilon -\delta A^2 +\frac{3}{2}A^4
\right) \Theta(b)~, \label{eq:eqstatA}
\end{eqnarray}
where  $b = |\sum_je^{i\theta_j}|/N $ is the bunching parameter and $\Theta$ is
the reciprocal function of $I_1(x)/I_0(x)$, $I_n$ being the
modified Bessel function of order~$n$. Let us note that
Eqs.~(\ref{eq:eqstatm}) and~(\ref{eq:eqstatA}) give the
microcanonical solution of Hamiltonian~(\ref{eq:Hamiltonien}). The
canonical solution of the same Hamiltonian in the context of
plasma physics was obtained in~\cite{Firpo00}. It turns out that the
two ensembles are equivalent, which was not granted a priori for
such a self-consistent system with infinite range
interactions~\cite{Houches02}. Let us remark that
Eqs.~(\ref{eq:eqstatm}) and~(\ref{eq:eqstatA}) were obtained in
Ref.~\cite{Gluckstern93} using several hypothesis, suggested only
by numerical simulations. Here, a statistical mechanics approach
gives a complete and self-consistent framework to justify their
derivation. In particular, let us emphasize that, contrary to the
previous approach, it is not necessary to choose a priori the
distribution~$f$, which is fully determined by the method of solution.

\begin{figure}[th]
\resizebox{0.4\textwidth}{!}{\includegraphics{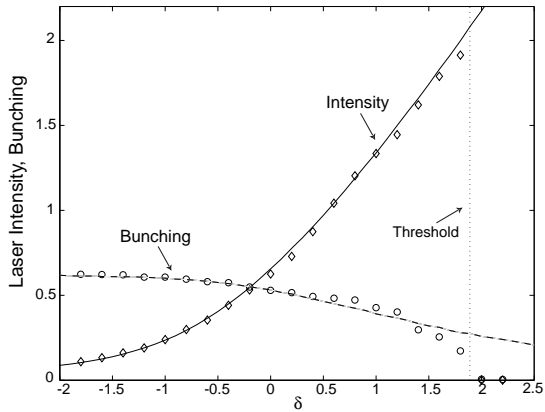}}
\caption{\label{fig:compasimul1} Comparison between theory (solid
and long-dashed lines) and simulations (symbols) for a monoenergetic beam,
varying the detuning $\delta$. The vertical dotted  line,
$\delta=\delta_c\simeq1.9$,  represents the transition from the low to
the high-gain regime.} \vskip -.5truecm
\end{figure}

Figure~\ref{fig:compasimul1} presents the comparison between the
analytical predictions and the numerical simulations performed using
equations (\ref{eq:mvttheta})-(\ref{eq:mvtA}) in the case of a
monoenergetic beam. Numerical data are time averaged. The agreement is
remarkably good for $\delta<0.5$, and is accurate up to the threshold value
$\delta_c$, although phase space mixing is probably less effective for
larger detuning. For $\delta>\delta_c$, there is no amplification, hence, both
intensity and bunching stick to their initial vanishing values.
This transition, purely dynamical, cannot be
reproduced by the statistical analysis.

\begin{figure}[th]

\resizebox{0.4\textwidth}{!}{\includegraphics{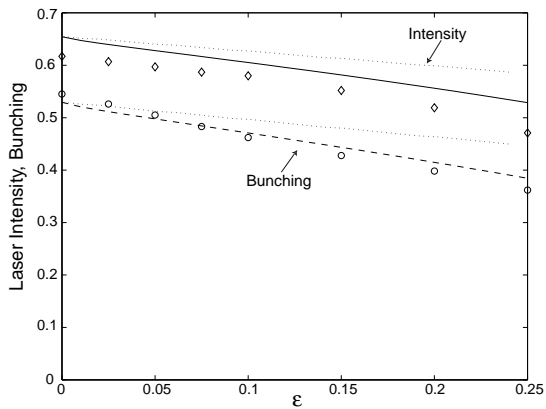}}
\caption{\label{fig:compasimul2} Comparison between theory (solid
and long-dashed lines) and simulations (symbols) for a non
monoenergetic beam when the energy $\varepsilon$,
characterizing the velocity dispersion of the initial
electron beam, is varied.  The dotted lines represent the
intensity and bunching predicted by the full statistical
equilibrium given by Eqs.~(\ref{eq:eqstatm},\ref{eq:eqstatA}), not
very appropriate here, whereas the solid line and long-dashed
lines refer to the Vlasov equilibrium defined by
Eqs.~(\ref{eq:barf}) and (\ref{eq:A}). The discrepancy between
theory and numerical experiments is small over the whole range of
explored energies.} \vskip -.5truecm
\end{figure}

In the case of a non monoenergetic beam,
Figure~\ref{fig:compasimul2} presents the comparison between the
estimates of the above theoretical analysis and the results of
direct numerical simulations of
Hamiltonian~(\ref{eq:Hamiltonien}), after time
averaging. The comparison is shown in the energy range that
allows the amplification process to take place. The good agreement
for intensity and bunching provides therefore an a posteriori,
but striking, support for Vlasov statistical equilibrium.

In this paper, we have proposed a new approach to
study the saturated state of the Compton Free Electron Laser,
based on a {\em statistical mechanics approach} in the framework
of Colson-Bonifacio's model~\cite{Bonifacio90}.
By drawing analogies with
the statistical theory of violent relaxation in astrophysics and
2D Euler turbulence, we have derived analytical estimates of the
saturated intensity and bunching. In addition to providing a deeper
insight into the physical behaviour of this system, the results of
our theory agree very well with numerical simulations. Due to its
intrinsic flexibility, it may be possible to
adapt the statistical approach to more complete models and complex
schemes, thus allowing a direct comparison between analytical
studies and experiments on real devices. Such a statistical 
approach could be used as a tool to define future
strategies aiming at {\em optimizing} FEL performance.

We would like to thank Y. Elskens, L. Giannessi and S. Reiche
for useful discussions


\begin{thebibliography}{99}
\bibitem{aciter} The case of a linear undulator can also be treated
    by rescaling $a_w$ as  $a_w/\sqrt{2}$ and setting $A_y=0$.
\end{thebibliography}
\end{document}